%% file: high_pt2.tex
\documentstyle[12pt,epsfig]{article} 
\newlength{\dinwidth}                       
\newlength{\dinmargin}                      
\input{abbreviations}

\begin{document}
\noindent
{\tt MPI-PhE/96-18}    \hfill     \\
{\tt hep-ex/yymmnn}               \\
{\tt October 1996}                  \\

\vspace*{1cm}
\begin{center}  \begin{Large} \begin{bf}
 High--\pt Particles in the Forward Region at HERA
\footnote{Contribution to the Workshop on ``Future Physics at HERA'',
  Hamburg 1996.} \\
  \end{bf}  \end{Large}
  \vspace*{5mm}
  \begin{large}
M. Kuhlen\\ 
  \end{large}
Max-Planck-Institut f\"ur Physik,
Werner-Heisenberg-Institut, \\
F\"ohringer Ring 6,
D-80805 M\"unchen,
Germany,
E-mail: kuhlen@desy.de\\
\end{center}
\begin{quotation}
\noindent
{\bf Abstract:}
In order to probe the dynamics of parton evolution 
in deep inelastic scattering
at small \xb,
high--\pt particles produced centrally in pseudorapidity are studied.
In the BFKL mechanism gluon radiation is expected to be more abundant
than for DGLAP evolution with strong ordering of the gluon
transverse momenta, leading to harder \pt spectra.
The proposed measurements require charged particle tracking
capability as much forward as possible in the HERA laboratory frame, 
for example with a Very Forward Silicon Tracker,
and high luminosity for detailed studies.
\end{quotation}

%

\begin{figure}[htb]
   \centering
   \vspace{-0.2cm}
   \epsfig{file=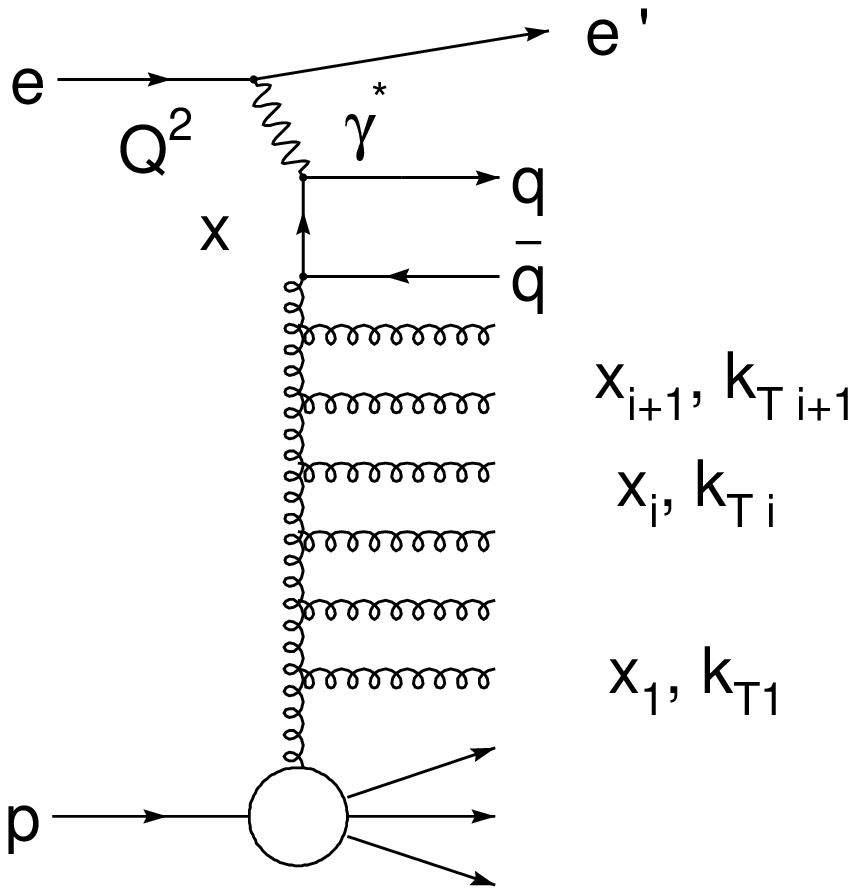,width=6cm}
   \epsfig{file=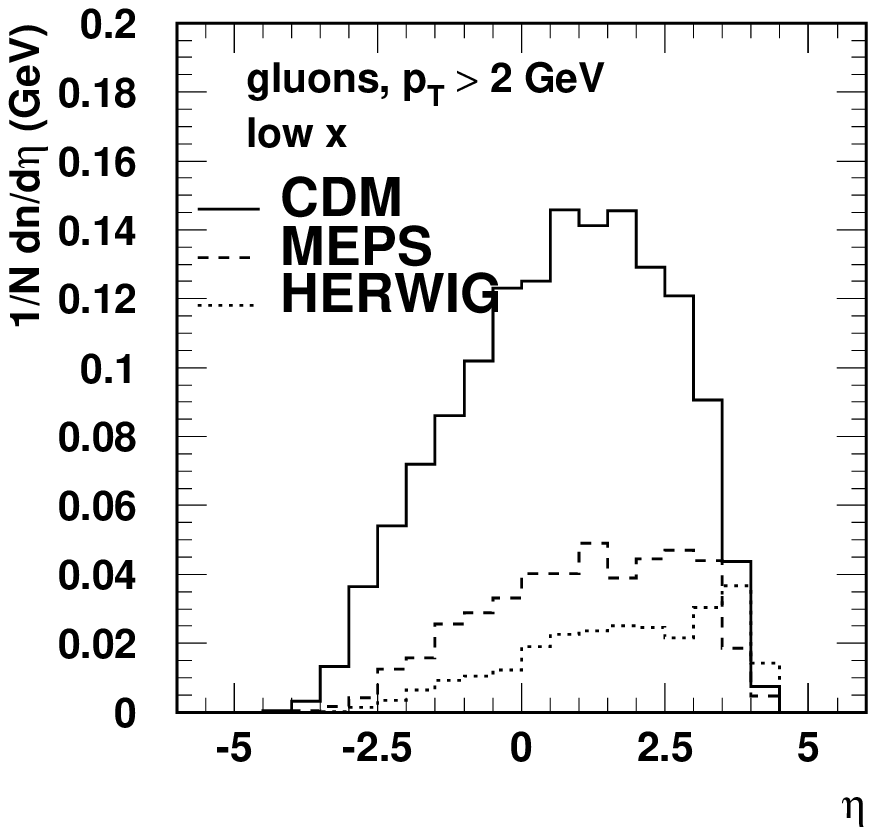,width=6cm,
    bbllx=4pt,bblly=275pt,bburx=257pt,bbury=516pt,clip=}
   \caption{\it 
       {\bf a)} Parton evolution cascade. 
       {\bf b)} The multiplicity of hard 
                gluons with $\pt>2\GeV$ as a function of $\eta$
                for events at small $x$,
                \av{$x$}=0.00037 at $\av{\Qsq}\approx 14 \GeVsq$. 
                The proton remnant direction is to the left.}
   \label{cascade} 
\end{figure}

HERA allows the study of a new kinematical regime in deep inelastic 
scattering, reaching very small values of Bjorken-\xb ($\approx 10^{-5}$)
with
\Qsq still a few \GeVsq, such that perturbation theory can
be applied. 
It is an open theoretical question whether HERA 
data can still be described with the conventional 
DGLAP 
(Dokshitzer-Gribov-Lipatov-Altarelli-Parisi) 
parton evolution 
equations \cite{dglap} which are derived for not too small \xb
and correspond to a resummation of 
terms proportional to $(\as \ln (\Qsq/ Q_0^2))$,
or whether terms  proportional to $(\as \ln 1/x)^n$ become
important, which are treated in the BFKL
(Balitsky-Fadin-Kuraev-Lipatov) 
equation \cite{bfkl}.
(More recently, 
in the CCFM approach \cite{ccfm} an equation for
both small and large \xb has been provided).
The two approximations lead to different constraints for the
gluons which can be emitted in the parton evolution chain
(Fig.~\ref{cascade}a). 
The leading log
DGLAP evolution corresponds to a strong
ordering of the transverse momenta \kt 
(with respect to the proton beam)
in the parton cascade
($Q_0^2 \ll \kt_1^2 \ll ... \kt_i^2 \ll ... \Qsq$),
while in the BFKL evolution arbitrary \kt are possible
\cite{ordering}.
It appears that the structure function measurements are 
too inclusive to distinguish the different evolutions \cite{martin}.
The hadronic final state 
emerging from the cascade may be more
sensitive to the new type of evolution (see e.g. \cite{kwiecinski}).
However, hadronization effects screen to some extent the parton dynamics
from direct observation \cite{kuhlen1}. 

QCD predictions for the 
hadronic final state
are extracted
from Monte Carlo models, which incorporate the QCD evolution in
different approximations and utilize
phenomenological models 
for the non-perturbative hadronization phase.
The MEPS (Matrix Element plus Parton Shower) \cite{lepto} 
and the HERWIG \cite{herwig} models invoke leading log DGLAP parton showers
with strong \kt ordering.
In the colour dipole model (CDM) \cite{dipole},
an unordered parton emission scenario is realized,
and in that respect it is similar to the BFKL evolution~\cite{bfklcdm}.
Gluon radiation in MEPS and HERWIG is
suppressed w.r.t. CDM without the \kt ordering constraint
(see Fig.~\ref{cascade}b)\footnote{All distributions 
shown are in the hadronic centre of mass
system (CMS), and are normalized to the number of events $N$ which enter
the distribution. The pseudorapidity $\eta$ in the CMS is defined as 
$\eta=-\ln\tan\theta/2$, with $\theta$ measured 
with respect to the virtual photon
direction.}.
However, all models give a satisfactory overall description of current HERA
final state data~\cite{carli}.

\begin{figure}[htb]
   \centering
   \vspace{-0.2cm}
   \epsfig{file=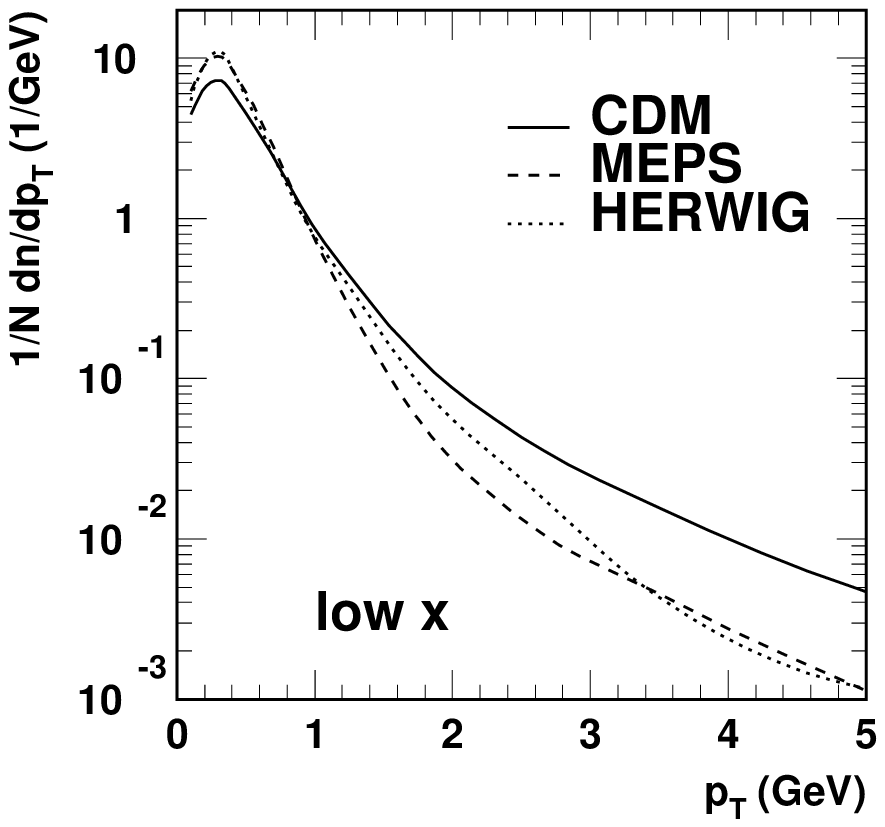,width=6cm,
    bbllx=4pt,bblly=275pt,bburx=257pt,bbury=516pt,clip=}
   \epsfig{file=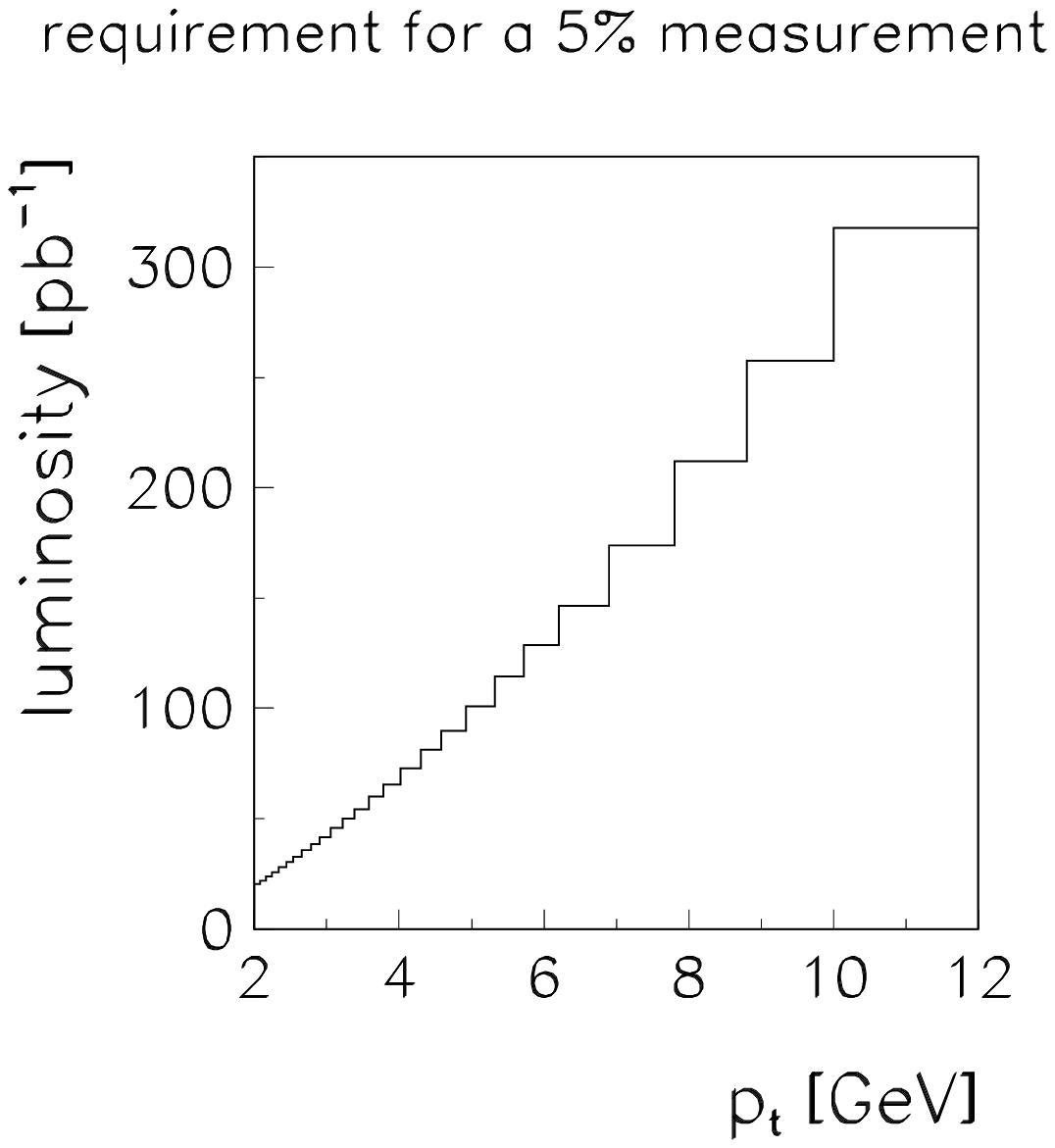,width=6cm,
    bbllx=-100pt,bblly=18pt,bburx=308pt,bbury=314pt,clip=}
   \caption{\it 
       {\bf a)} The \pt spectra for charged hadrons from $0<\eta<2$
                at \av{$x$}=0.00037 and $\av{Q^2}\approx 14 \GeVsq$.
       {\bf b)} The luminosity needed for a statistical precision of 5\%
       for a given bin in $p_T$.} 
   \label{lumi} 
\end{figure}

%
During the HERA workshop
it was found that
inclusive charged particle transverse momentum 
($p_T$) spectra offer a handle 
to disentangle hard perturbative 
from soft hadronization effects \cite{kuhlen2}. 
The hard tail of the \pt spectra (see Fig.~\ref{lumi}a)
is sensitive to
parton radiation from the cascade.
The CDM generates a harder tail than LEPTO and HERWIG
with suppressed gluon radiation.
One can thus hope to study the parton evolution dynamics with high--\pt
particles, and discriminate between the different scenarios.
Here
the implications for
the future experimentation at HERA will be investigated.

The measurement of the hard tail of the \pt spectrum
shown in Fig.~\ref{lumi}a would pose a challenge to QCD
(in fact, a QCD calculation for the rate of high \pt forward pions,
based upon the BFKL equation, has just become available
\cite{klm}), 
and it has been investigated what experimental precision can be achieved.
In Fig.~\ref{lumi}b the luminosity needed is shown 
to reduce the statistical error for a given \pt bin to the same
level as the expected systematic error
\cite{wieland} for that bin. 
The luminosity estimate is based upon the CDM.
The binning was chosen such that
the systematic error is constant ($\approx 5\%$) for all bins.
In order to measure the \pt spectrum at \pt=12\GeV with a 
precision of 5\% for both statistical and systematic error
an integrated luminosity of 300~\pbinv would be needed. 
The quest for a measurement at large \pt (and large luminosity)
is also motivated by the expectation that perturbative
QCD will be more reliable there.

\begin{figure}[htb]
   \centering
   \vspace{-0.2cm}
   \epsfig{file=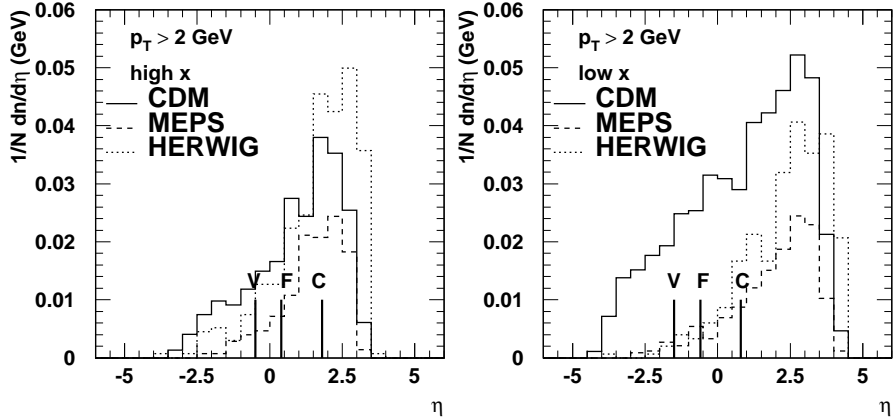,width=12cm,
    bbllx=2pt,bblly=276pt,bburx=518pt,bbury=521pt,clip=}
   \caption{\it 
       Multiplicity flow vs. $\eta$
       for charged hadrons with $\pt > 2 \GeV$ at
       {\bf a)} ``high'' \xb and
       {\bf b)} ``low'' \xb.
       The current direction is to the right.
       Indicated are the acceptance limits of the H1 central tracking
       system (C, $\thlab > 20 \degr$), 
       the forward tracking system (F, $\thlab > 7 \degr$), and a 
       hypothetical very forward tracking system (V, $\thlab > 3 \degr$).  
       Here \thlab is measured in the laboratory system w.r.t. the 
       proton direction.} 
   \label{dndeta} 
\end{figure}

As a measure of the rate of hard particles as a function
of pseudorapidity, the multiplicity flow of charged particles
with $\pt > 2 \GeV$ is shown vs. $\eta$
in Fig.~\ref{dndeta}. 
Events from two kinematic bins, 
one at ``large \xb'' (\av{$x$}=0.0023) and one at ``small $x$'' 
(\av{$x$}=0.00037) are compared, with $\av{Q^2}\approx 14 \GeVsq$ 
approximately constant. 
At high \xb the differences between the model predictions 
are not very big,
but at small \xb the models deviate by a large amount. The CDM
produces much more particles with $\pt > 2 \GeV$ than MEPS and 
HERWIG, and the discrepancy increases with the distance from the
current system.
That difference has its origin in the very different gluon
emission pattern, see
Fig.~\ref{cascade}b. 

The sensitivity to this effect with typical (here H1 \cite{h1nim}) HERA 
tracking devices is indicated. It is clear that such 
a measurement requires charged particle tracking capability as far
``forward'' (into the remnant direction) as possible. 
The so called ``forward'' region of the HERA detectors corresponds
to the central region in the hadronic CMS.
Rather than measuring with high precision a relatively small effect,
which may be masked by hadronization and other uncertainties,
it is preferable to measure a large effect
with moderate precision. 
Therefore a ``Very Forward Tracker''
is proposed which covers angles down to $\thlab=3\degr$.

Here a feasibility study for the case of H1 is presented.
A silicon tracker 
similar to the H1 Backward Silicon Tracker (BST), 
but with radial (``$\phi$'') readout strips to measure curvature,
could be positioned 
suitably in the forward region \cite{bst}.
It would
consist of four to eight disks mounted perpendicular to the beam line,
and sit in between the central drift chamber and the beam pipe.
There is space available between $z\approx 40-120$~cm, with
$z$ being the longitudinal distance from the interaction point. 
Radially,
the sensitivity would extend from 6 to 12 cm.
Angular coverage from $\thlab=3\degr$ to  $8\degr$ would be possible,
matching with the end of the H1 forward tracker acceptance at $\thlab=7\degr$.
With four readout planes, 
spaced 0.1 m apart, 
and with a pitch of 50 $\mu$m, a momentum
resolution 
of $\delta \pt / \pt \approx 10 \% \cdot \pt$ can be achieved.
It can be improved by a vertex constraint. 
Since the effective HERA beam width 
is 70 $\mu$m (vertical) by  330 $\mu$m (horizontal),
the event vertex needs to be defined by tracks measured in the 
existing central
silicon tracker \cite{bst} (impact parameter resolution 60 $\mu$m), 
reducing it to roughly 50~$\mu$m by 50~$\mu$m.
When such a vertex constraint is applied, the momentum resolution
could be improved to $\delta \pt / \pt \approx 2-3 \% \cdot \pt$.
That is certainly sufficient for the simple measurement of the \pt
spectra discussed here. 
More detailed studies and simulations would be neccessary to study
the occupancy in the detector and questions of pattern recognition.
The detector would cost 250k-500k DM (estimate based upon the BST costs), 
depending on the number of silicon planes.

The HERA luminosity upgrade would have an impact on the VFT. 
When dipoles are inserted into
the H1 detector,
perhaps up to the faces of the central drift chamber,
that would require a larger diameter beam pipe to let out the produced
synchrotron radiation.
The existing BST could no longer be used, and the VFT would 
need to be redesigned according to the larger diameter beam pipe.
Of course other options could also be explored.
For example, H1 is building a 
Very Low Q-square (VLQ) tracker for the backward region
based upon GaAs technology. 
Again, this device would not fit on a larger diameter beam pipe.
However, it may be possible to save on the new detector by
re-using the existing electronics.

Initially,
for a minimal meaningful measurement a moderate integrated luminosity
of $10 \pbinv$ would already be sufficient.
For that purpose, one could even think of a short dedicated
HERA run with minimal, but optimized instrumentation, including only
the trackers and backward electron detection devices.

In this study, however, only the sensitivity
of charged particle 
spectra\footnote{
Other particles could be utilized as well.
For example,
$\pi^0\rightarrow\gamma\gamma$ could be detected in the forward calorimeter,
or $K^0_S$ decays could be identified via secondary vertices.
It may even be possible to identify photons either in the calorimeter
or via conversions seen in the trackers.}
to suppressed or abundant gluon
radiation scenarios has been exploited. It should be possible
to construct variables based on correlations between high--\pt
particles which probe the gluon dynamics,
for example \kt ordering or recombination effects,  
more directly and locally. 
Considering
the rate of such particles (Fig.~\ref{dndeta}), 
high luminosity would
be needed to have enough events with pairs of high--\pt particles.  
Again, a large forward acceptance would increase the evolution length
that can be probed,  
as well as the statistics of high--\pt particle pairs
for correlation studies.

\end{document}

%% file: abbreviations.tex
\newcommand{\av}[1]{\mbox{$ \langle #1 \rangle $}}
\def\lsim{\mathrel{\rlap{\lower4pt\hbox{\hskip1pt$\sim$}}
    \raise1pt\hbox{$<$}}}                
\def\gsim{\mathrel{\rlap{\lower4pt\hbox{\hskip1pt$\sim$}}
    \raise1pt\hbox{$>$}}}                

\newcommand{\xb}{\mbox{$x~$}}  

\newcommand{\Qsq}{\mbox{$Q^2~$}}

\newcommand{\kt}{\mbox{$k_T~$}}
\newcommand{\pt}{\mbox{$p_T~$}}

\newcommand{\thlab}{\mbox{$\theta_{\rm lab}~$}}

\newcommand{\as}{\mbox{$\alpha_s~$}}

\newcommand{\degr}{\mbox{$^\circ$}}

\newcommand{\GeV}{\mbox{\rm ~GeV~}}

\newcommand{\GeVsq}{\mbox{${\rm ~GeV}^2~$}}

\newcommand{\pbinv}{\mbox{${\rm ~pb^{-1}}~$}}


%% file: high_pt2.bbl
\begin{thebibliography}{99}

\bibitem{dglap}
  Yu. L. Dokshitzer, Sov. Phys. JETP 46 (1977) 641; \\
  V.N. Gribov and L.N. Lipatov, Sov. J. Nucl. Phys. 15 (1972) 438 and 675; \\
  G. Altarelli and G. Parisi, Nucl. Phys. 126 (1977) 297.

\bibitem{bfkl}
  E.A. Kuraev, L.N. Lipatov and V.S. Fadin, Sov. Phys. JETP 45 (1972) 199; \\
  Y.Y. Balitsky and L.N. Lipatov, Sov. J. Nucl. Phys. 28 (1978) 282. 

\bibitem{ccfm}
  M. Ciafaloni, Nucl. Phys. B296) (1988) 49; \\ 
  S. Catani, F. Fiorani and G. Marchesini, Phys. Lett. B234 (1990) 339;
  Nucl. Phys. B336 (1990) 18.

\bibitem{ordering}
  J. Bartels, H. Lotter,  Phys. Lett. B309 (1993) 400; \\
  J. Bartels, H. Lotter and M. Vogt, DESY-95-224.

\bibitem{martin}
  A.D. Martin, to appear in 
  Proc. of the Workshop on Deep Inelastic Scattering 
  and Related Phenomena, DIS96, Rome 1996.

\bibitem{kwiecinski}
  J. Kwieci\'{n}ski, Nucl. Phys. B, Proc. Suppl. 39BC (1995) 58.

\bibitem{kuhlen1}
  M. Kuhlen, 
  hep-ex/9508014; 
  Proc. of the Workshop on Deep Inelastic Scattering and QCD - DIS95,
  Paris, April 1995, eds. JF. Laporte and Y. Sirois, p. 345.









\bibitem{lepto} 
  G. Ingelman,       
  Proc. of the Workshop on Physics at HERA, Hamburg 1991,
  eds. W. Buchm\"uller and G. Ingelman, 
  vol. 3, p. 1366; \\
  A. Edin, G. Ingelman and J. Rathsman, Phys.Lett. B366 (1996) 371.


\bibitem{dipole}
 G. Gustafson, Ulf Petterson, Nucl. Phys. B306 (1988); \\
 G. Gustafson, Phys. Lett. B175 (1986) 453; \\
 B. Andersson, G. Gustafson, L. L\"onnblad, Ulf Petterson, 
 Z. Phys. C43 (1989) 625.

\bibitem{ariadne}  
  L. L\"onnblad,
  Computer Phys. Comm. 71 (1992) 15.





\bibitem{herwig}
  G. Marchesini et al., 
  Computer Phys. Comm. 67 (1992) 465.



\bibitem{bfklcdm}
 L. L\"onnblad, Z. Phys. C65 (1995) 285; CERN-TH/95-95;\\
 A. H. Mueller, Nucl. Phys. B415 (1994) 373.

\bibitem{carli}
  T. Carli, to appear in
  Proc. of the Workshop on Deep Inelastic Scattering 
  and Related Phenomena, DIS96, Rome 1996.

\bibitem{kuhlen2}
  M. Kuhlen, Phys. Lett. B382 (1996) 441.

\bibitem{klm} J. Kwieci\'{n}ski, S.C. Lang and A.D. Martin, 
 Durham preprint DTP/96/62.

\bibitem{wieland}
  W. Hoprich, diploma thesis, Univ. Heidelberg, 1996.

\bibitem{h1nim}
  H1 Collab., I. Abt et al., 
  DESY 93-103 (1993).

\bibitem{bst}
  N. Wulff, H1 internal note H1-09/91-191; \\
  J. B\"urger et al., Nucl. Instr. Meth. A367 (1995) 422; \\
  P. Biddulph and D. Pitzl, private communication.




\end{thebibliography}
